\documentclass[12pt,tightenlines,eqsecnum,floats,aps,amsmath,amssymb,nofootinbib,prd,superscriptaddress]{revtex4}

\usepackage{tikz-feynman, contour}
\usepackage{feynmp-auto}

\usepackage{hyperref}
\usepackage{comment}
\usepackage{color}

\usepackage{setspace}
\usepackage{amsmath,amssymb,amsfonts,amsthm,mathrsfs}
\usepackage{cancel}
\usepackage{graphicx,float}
\usepackage[export]{adjustbox}
\usepackage{enumerate} 
\usepackage{slashed}

\def\ba{\begin{eqnarray}}
\def\ea{\end{eqnarray}}
\def\bi{\begin{itemize}}
\def\ei{\end{itemize}}

\newcommand{\be}{\begin{equation}}
\newcommand{\ee}{\end{equation}}

\def\D{\mathcal{D}}

\def\A{\mathcal{A}}
\def\F{\mathcal{F}}
\def\I{\mathcal{I}}
\newcommand{\ov}[2]{\overset{#1}{#2}}
\def\e{\varepsilon}

\def\hard{\textrm{hard}}
\def\soft{\textrm{soft}}

\def\w{\omega}
\def\zb{\bar{z}}

\def\L{\mathcal{L}}
\def\J{\mathcal{J}}
\def\Psib{\overline{\Psi}}

\def\dpp{\widetilde{d p}\,}
\def\Yb{\bar{Y}}
\def\psit{\tilde{\psi}}
\def\At{\tilde{A}}

\def\Ar{\text{A}}
\def\Arb{\text{\bf A}}
\def\ub{\bar{u}}
\def\eb{\bar{e}}
\def\psib{\bar{\psi}}
\def\psibt{\tilde{\bar{\psi}}}
\def\chib{\bar{\chi}}
\def\hb{\bar{h}}
\def\lps{\text{LPS}}
\def\ubold{\mathbf{u}}

\begin{document}
\title{Fermionic asymptotic symmetries in massless QED}
\author{Adri\'an Agriela}
\email{aagriela@fing.edu.uy}
\affiliation{Instituto de Física, Facultad de Ciencias, Universidad de la Rep\'ublica \\ Igu\'a 4225, Montevideo, Uruguay}
\affiliation{Instituto de Física, Facultad de Ingeniería, Universidad de la Rep\'ublica \\ Julio Herrera y Reissig 565, Montevideo, Uruguay}
\author{Miguel Campiglia}
\email{campi@fisica.edu.uy} 
\affiliation{Instituto de Física, Facultad de Ciencias, Universidad de la Rep\'ublica \\ Igu\'a 4225, Montevideo, Uruguay}

\begin{abstract}

We consider soft electrons in massless QED at tree-level. The emission amplitude at leading order in the soft electron energy factorizes in a way similar to the soft photon case. We recast the soft electron factorization formula as a Ward identity of an asymptotic charge. This leads to the first example of an asymptotic fermionic symmetry in a theory with no conventional supersymmetry, suggesting that tree-level massless QED may posses an asymptotic supersymmetry algebra.  
Although our  approach  does not yet allow us to completely characterize the algebra, it suggests that subleading soft photons should feature in the 
 anticommutator of two fermionic symmetry generators.




\end{abstract}

\maketitle

\tableofcontents

\section{Introduction}
Asymptotic states of massless particles have their natural home at null infinity \cite{penrose}. At this boundary of spacetime it is possible to unveil  
symmetries that are otherwise obscure from a bulk perspective  \cite{asymquant,stromlectures}. As first shown by Strominger and collaborators \cite{stromST,stromVir,stromQED}, these asymptotic symmetries manifest themselves in scattering amplitudes via so-called soft theorems. Conversely, soft theorems in scattering amplitudes can often be interpreted as arising from asymptotic symmetries, see e.g. \cite{stromLow,stromfermionic,lysovgravitino,averygravitino,subsub}. In this article we present what may be the simplest example of an asymptotic \emph{fermionic} symmetry. We consider tree-level massless Quantum Electrodynamics (QED), in which there is a notion of soft electrons, i.e. electrons with vanishingly small energy.\footnote{Alternatively, we are  studying massive electrons in a regime where the soft energy $E_\soft$ is much larger than the electron mass yet much smaller than all other energies involved in the process:   $m_e \ll E_\soft  \ll E_\hard  $.} The emission amplitude for such  soft electrons is given to leading order by a soft electron theorem.  We will recast such theorem as a Ward identity of fermionic asymptotic charges,  thus suggesting  the existence of  asymptotic fermionic symmetries.  At this stage, however, we are  unable to fully characterize 
 the underlying symmetry algebra.  In particular, we lack a trustworthy evaluation of the commutator between fermionic symmetries. We will describe what the obstacles are and discuss possible strategies to overcome them.


Fermionic asymptotic symmetries have already been discussed in other contexts. They naturally arise in supergravity theories \cite{Awada:1985by,Henneaux:2020ekh,Fuentealba:2020aax,Fuentealba:2021xhn,banerjee} where they are associated to soft gravitinos \cite{lysovgravitino,averygravitino}. In \cite{stromfermionic}, Dumitrescu, He, Mitra and Strominger identified asymptotic fermionic charges in supersymmetric abelian gauge theories. Our work follows closely  their  analysis, with the electron field here playing the role of photino field there. A comprehensive discussion of (conformally) soft fermions can be found in  \cite{Pano:2021ewd}.

The organization of the paper is as follows. In the next section we introduce notation and review basic concepts on radiative phase spaces and asymptotic symmetries, in the context of massless QED.  In section \ref{softelectron} we present a soft electron theorem and interpret it as a Ward identify of an asymptotic fermionic charge. In section \ref{fermionicsymsec} we discuss various aspects of the associated fermionic symmetries, including the commutation relations with  bosonic symmetries. We conclude in section \ref{outlook}, where we highlight the open questions  left for future work. Additional material is given in three appendices: In appendix \ref{PBapp} we review the relationship  between the momentum space and null infinity descriptions of fields. In appendix \ref{softparticlesapp} we review the soft photon theorem and derive the analogous soft electron theorem. In appendix \ref{subphapp} we present a preliminary exploration on the non-linear structure of the fermionic symmetry, and observe how  a naive evaluation of the commutator of two fermionic symmetries displays  similarities with the asymptotic symmetry associated to subleading soft photons.




\section{Preliminaries} \label{prelsec}
\subsection{Conventions}
The elementary field variables for QED are the $U(1)$ gauge field $\A_\mu$ and the anticommuting Dirac spinor $\Psi$. In the massless case the lagrangian density is\footnote{We follow conventions from \cite{peskin} modulo a sign in the definition of the coupling constant $e$.} 
\be \label{L}
\L= \sqrt{-\eta}\left(- \frac{1}{4} \F^{\mu \nu} \F_{\mu \nu}+  i \Psib  \gamma^\mu \D_\mu \Psi \right),
\ee
where $\sqrt{-\eta}$ is the Minkowski  volume element, $\F_{\mu\nu}= \partial_\mu \A_\nu - \partial_\nu \A_\mu$ is the field strength, $\gamma^\mu$ Dirac matrices, $\Psib=\Psi^\dagger\gamma^0 $ and
\be
\D_\nu = \partial_\mu + i e \A_\mu
\ee
the gauge covariant derivative.  The electric current is defined as
\be \label{defJ}
 \J^\mu = e \Psib \gamma^\mu  \Psi.
\ee
Taking variations of the lagrangian density, one finds
\be
\delta \L = eom + \partial_\mu \theta^\mu(\delta)
\ee
 where
 \be
 eom = (\partial_\mu \F^{\mu \nu} - \J^\nu) \delta \A_\nu + (i \delta \Psib  \gamma^\mu \D_\mu \Psi + c.c.)
 \ee
yield the field equations  and
\be \label{defsymcurr}
\theta^\mu(\delta) = \sqrt{-\eta}(-\F^{\mu \nu}\delta \A_\nu + i \Psib \gamma^\mu \delta \Psi)
\ee
is the symplectic potential current. 

Besides Poincare (and in fact Conformal) symmetries, the theory is invariant under local gauge transformations
\be \label{delLambda}
\delta_\Lambda \A_\mu = \partial_\mu \Lambda , \quad \delta_\Lambda \Psi = - i e \Lambda \Psi,
\ee
as well as global axial rotations
\be \label{deltaA}
\delta_{\Ar} \A_\mu =0, \quad \delta_{\Ar} \Psi =  i \gamma^5 \Psi.
\ee
The latter famously displays a 1-loop anomaly \cite{abj}, but for the purposes of our  tree-level  discussion, we will regard \eqref{deltaA} as an exact symmetry. 

\subsection{Asymptotic fields at null infinity}

\subsubsection{Outgoing null coordinates}
In order to describe the fields near (future) null infinity, it is convenient to work in outgoing null coordinates. These can be defined as follows. First, assign a future null direction $q^\mu$ to every point  $x \equiv (z,\zb)$ on the celestial sphere,
\be \label{defq}
q^\mu(x) := \frac{1}{\sqrt{2}}\left( 1+ |z|^2, z+ \zb, -i (z- \zb), 1- |z|^2 \right),
\ee
The specific choice \eqref{defq} leads to a flat conformal frame on the celestial sphere, see e.g. \cite{Kapec:2017gsg} for a discussion on other possible frames.  Next,  choose a reference null vector $k^\mu$ transverse to $q^\mu$ that specifies the ``flow of time'',
\be \label{defk}
k^\mu = \frac{1}{\sqrt{2}} \left(1,0,0,-1 \right), \quad k^\mu q_\mu(x) =-1.
\ee
Finally,  parametrize cartesian coordinates  $X^\mu$ by  $(r,u, x)$ according to     
\be \label{Xitorux}
X^\mu(r,u,x) = r q^\mu(x) + u k^\mu,
\ee
in terms of which the  spacetime metric takes the form 
\be
d X^\mu d X_\mu =  -2 du dr + 2 r^2 d z d \zb.
\ee

Future null infinity $\I$ is reached by taking $r \to \infty$ with constant $(u,x)$. One can similarly define retarded null coordinates that are adapted to past null infinity. We will however focus our discussion on fields at future null infinity, with the understanding that a parallel construction is available at past null infinity.
\subsubsection{Fall-offs in $r$}
Near null infinity the electromagnetic field is described by the leading transversal components  \cite{asymquant,stromlectures},
\be
\A_z(r,u,x) \stackrel{r \to \infty}{=} A_z(u,x) + \cdots, 
\ee
where $A_z$ is regarded as a gauge field on $\I$. The massless Dirac field fall-offs have been discussed in \cite{stromfermionic}. As for other massless fields, it decays as the inverse power of the radial coordinate.\footnote{We are assuming free-field fall-offs, which suffice for the tree-level considerations of this work. Loop corrections may imply slower fall-offs, see e.g.  \cite{sayali}.} The Dirac equation  imposes restrictions on the leading spinor components, leaving only two independent asymptotic fields:\footnote{Further details are given in the discussion following  Eq. \eqref{Diraceqinr}.}
\be \label{asymPsi}
\Psi \stackrel{r \to \infty}{=} \frac{1}{ 2^{1/4} r } \begin{pmatrix} -\zb \psi_- \\ \psi_- \\ \psi_+ \\ z \psi_+  \end{pmatrix} + \cdots ,
\ee
where the overall normalization is chosen for later convenience and  $\psi_\pm$ are regarded as complex fermionic fields on $\I$. 
Upon quantization, they   describe  electrons of positive/negative helicity. Similarly, $A_z$/$A_{\zb}$ describe photons of  positive/negative helicity.  We shall  use the notation 
\be \label{Apm}
A_+ := A_z, \quad A_- := A_{\zb}.
\ee
Notice that, unlike $\psi_\pm$, the gauge field satisfies the reality condition  $A^*_+=A_-$.

\subsection{Radiative phase space}
Given the fall-offs from the previous section, we can evaluate the symplectic potential current \eqref{defsymcurr} at null infinity. For the relevant component $\mu=r$ one finds
\be \label{thetascri}
\lim_{r \to \infty}\theta^r(\delta)  = \dot{A}_z \delta A_{\zb} + \dot{A}_{\zb} \delta A_{z} + i  \psib_+ \delta \psi_+ + i \psib_- \delta \psi_-
\ee
where $\dot{A}_z\equiv \partial_u A_z$ and $\psib_{\pm}$ is the complex conjugate of $\psi_{\pm}$.  Taking a second variation in \eqref{thetascri} and integrating over $(u,z,\zb)$ we obtain the symplectic structure at $\I$ \cite{AS},
\be \label{defOmega}
\Omega := \sum_{s=\pm} \int_{\I} du d^2 x  \left( \delta \dot{A}_{-s} \wedge \delta A_{s} + i \delta \psib_s \wedge \delta \psi_s \right),
\ee
where we used the notation \eqref{Apm} for the asymptotic gauge field. Evaluating \eqref{defOmega} on two variations $\delta_1$ and $\delta_2$ one has
\be \label{Omega12}
\Omega(\delta_1,\delta_2) = \sum_{s} \int_{\I}  \left( \delta_1 \dot{A}_{-s} \delta_2  A_{s} + i \delta_1 \psib_s \delta_2 \psi_s \right) - (\delta_1 \leftrightarrow \delta_2).
\ee
In the above expressions it is important that  $\psi_\pm$ are regarded as anticommuting  fields. In particular,  the reality of the symplectic form follows from the property $\overline{\psi_1 \psi_2}=\psib_2 \psib_1$. 

\subsubsection{Fall-offs in $u$} \label{fallusec}
In order for \eqref{defOmega} to be well-defined, we must impose fall-offs conditions on the fields as $|u| \to \infty$. We shall henceforth assume them to be
\ba
A_s(u,x) & \stackrel{|u|\to \infty}{=} &  O(1) + O(1/|u|^{\epsilon}) \label{falluA} \\
\psi_s(u,x) & \stackrel{|u|\to \infty}{=} & O(1/|u|^{1+\epsilon}) \label{fallupsi} 
\ea
for some  $\epsilon>0$.  Condition \eqref{falluA} is slightly more general than the typical scattering fall-off,  which corresponds to $\epsilon=1$ \cite{laddhasen}. Condition \eqref{fallupsi} is stronger than a minimal requirement for convergence of \eqref{defOmega} (for which it would be enough a fall-off faster than $1/|u|^{1/2}$). We require \eqref{fallupsi} to ensure finiteness of the asymptotic fermionic charge defined in section \ref{softelectron}.

\subsubsection{Radiative PBs}
From the symplectic  structure \eqref{defOmega} one can obtain the elementary non-trivial Poisson brackets (PBs)\footnote{Our conventions are as follows. The PBs between two functions $F$ and $G$ is given by $\{F,G\} := X_G(F)=-X_F(G)= \Omega(X_G,X_F )$ where $X_F$ is 
defined by the condition $\Omega(\delta,X_F ) = \delta F$ (and similarly for $X_G$). If $F$ and $G$ are fermionic, there are additional signs that can be determined by requiring the grasmannian Leibinitz rule on PBs, namely $\{F_1 F_2,G \} = F_1 \{F_2,G \} \pm \{F_1,G\} F_2 $ where the minus occurs if both $F_2$ and $G$ are fermionic. In particular the PBs between two fermionic functions is symmetric rather than antisymmetric. We refer to chapter 6 of \cite{HT} for further details on fermionic PBs. \label{hvfnote}  }
\be \label{elemPBs}
\begin{aligned}
\{A_s(u,x), \dot{A}_{-s}(u',x') \} & =& \frac{1}{2} \delta(u-u') \delta^{(2)}(x,x') ,\\
\{\psi_s(u,x), \psib_{s}(u',x') \} & =& - i \delta(u-u') \delta^{(2)}(x,x').
\end{aligned}
\ee

We recall however a well  known subtlety with the  gauge field PBs     \cite{stromQED}, which is the occurrence of a $1/2$ discontinuity that can be expressed as
\be \label{discontinuity}
\int^{\infty}_{-\infty} du \{ \cdot,   \dot{A}_s(u,x) \}  = \frac{1}{2} \{ \cdot, \int^{\infty}_{-\infty} du  \dot{A}_s (u,x)\}  ,
\ee
where $\{ \cdot , F\} \equiv X_F$ is the Hamiltonian vector field of a functional $F$ (see footnote \ref{hvfnote}). A fix to this problem was proposed in \cite{stromQED} via the isolation of the zero mode component of the gauge field and the introduction of boundary terms in the symplectic structure. For simplicity 
we will continue to work with the standard radiative phase space symplectic structure while   keeping care when needed of the aforementioned subtlety. 

\subsection{Asymptotic Fock space} \label{focksec}
To obtain the asymptotic Fock space of photons and massless electrons, we start by considering the  Fourier transform of the fields with respect to the $u$-variable,
\be \label{fouriernull}
\begin{aligned}
A_s(u,x) = \int_{-\infty}^\infty \frac{d \w}{2 \pi} \tilde{A}_s(\w,x) e^{-i \w u} , \\
 \psi_s(u,x) = \int_{-\infty}^\infty \frac{d \w}{2 \pi} \tilde{\psi}_s(\w,x) e^{-i \w u}, \\
 \psib_s(u,x) = \int_{-\infty}^\infty \frac{d \w}{2 \pi} \psibt_s(\w,x) e^{i \w u}.  \\
\end{aligned}
\ee
The brackets \eqref{elemPBs} imply
\be \label{omegaPBs}
\begin{aligned}
\{\At_s(\w,x), \At_{-s}(\w',x') \} & =& \frac{i \pi}{\w'} \delta(\w+ \w') \delta^{(2)}(x,x') ,\\
\{\psit_s(\w,x) , \psibt_{s}(\w',x') \} & =& - i 2 \pi \delta(\w-\w') \delta^{(2)}(x,x').
\end{aligned}
\ee
Notice that the information of $A_s(u,x)$ is contained in $\tilde{A}_s(\w,x), \w>0$ since $\tilde{A}_s(-\w,x)=(\tilde{A}_{-s}(\w,x))^*$ due to the reality of the gauge field. On the contrary, there is no relation between the positive and negative frequency components of $\tilde{\psi}_s(\w,x)$. An independent set of mode functions is then given by:
\be \label{fockops}
\begin{aligned}
a_s(\w,x) & := & \At_s(\w,x) &, \quad \w>0 \\
b_s(\w,x) & := & \psit_s(\w,x) &, \quad \w>0 \\
c_s(\w,x) & := & \psibt_{-s}(-\w,x)&, \quad  \w>0 .
\end{aligned}  
\ee
Upon quantization, these become the annihilation Fock operators of photons, electrons and positrons respectively.\footnote{The normalization however is different from the standard momentum-space Fock operators, see appendix \ref{PBapp}.} The (anti) commutation relations of these operators with their hermitian adjoints  are  dictated by $i$ times the PBs \eqref{omegaPBs}, from which one finds
\be \label{elemcomms}
\begin{aligned}
{[ a_s(\w,x) , a^\dagger_s(\w',x') ]} & =\frac{\pi}{\w} \delta(\w - \w') \delta^{(2)}(x,x'), \\
[ b_s(\w,x) , b^\dagger_s(\w',x') ] &= [ c_s(\w,x) , c^\dagger_s(\w',x') ] =2 \pi \delta(\w - \w') \delta^{(2)}(x,x')  .
\end{aligned}
\ee

It will be useful for later purposes to define  angular-density operators for each kind of particle:
\ba
\rho^A_s(x) & := &  \int_0^\infty \frac{d \w}{ \pi}  \w  a^\dagger_s(\w,x)  a_s(\w,x)  \\
\rho^\psi_s(x) & := & \int_0^\infty \frac{d \w}{2 \pi} b^\dagger_s(\w,x)  b_s(\w,x)  \\
\rho^{\bar{\psi}}_s(x) & := & \int_0^\infty \frac{d \w}{2 \pi} c^\dagger_s(\w,x)  c_s(\w,x)  
\ea
as well as particle-number operators
\be \label{defNX}
N^X_s  :=   \int d^2 x \rho^X_s(x) , \quad  X =\{A, \psi, \bar{\psi} \}.
\ee
We finally note the identity 
\be
 \int_{-\infty}^\infty du \psib_s(u,x) \psi_s(u,x)  =  \rho^\psi_s(x) - \rho^{\bar{\psi}}_{-s}(x).
\ee

\subsection{Bosonic (asymptotic) symmetries} \label{bossymsec}

In this section we review  bosonic symmetries of tree-level massless QED at null infinity: Poincare, large $U(1)$ gauge and axial rotations. 
We leave out of the discussion other bosonic symmetries that are more challenging to describe at null infinity:  Those arising from sub$^n$-leading soft photons\footnote{See however appendix \ref{subphapp} for a preliminary incorporation of subleading soft photon symmetries.} \cite{hamada,Li:2018gnc} as well as 4-dimensional conformal symmetry.

\subsubsection{Lorentz} 
Infinitesimal Lorentz transformations (or more generally superrotations) near null infinity are parametrized by holomorphic vector fields $Y(z) \partial_z$ (and their complex conjugate) by
\be \label{xiY}
\xi_Y = Y(z) \partial_z + \frac{1}{2}Y'(z)(u \partial_u - r \partial_r) + \cdots,
\ee
They act on fields via standard Lie derivatives, which for spinors include an internal rotation (see e.g. \cite{stromfermionic})
\ba \label{Liespinor}
\delta_\xi \Psi = \left(\xi^\mu \partial_\mu - \frac{1}{8} \nabla_{\mu} \xi_{\nu} [\gamma^\mu,\gamma^\nu] \right) \Psi.
\ea
Evaluating \eqref{Liespinor} for $\xi=\xi_Y$ on  \eqref{asymPsi} one finds
\be
\delta_{\xi_Y} \Psi = \frac{1}{ 2^{1/4} r } \begin{pmatrix} -\zb \delta_Y \psi_- \\ \delta_Y \psi_- \\ \delta_Y \psi_+ \\ z \delta_Y \psi_+  \end{pmatrix} + \cdots
\ee
where
\ba
\delta_Y \psi_- & = &  \left( Y(z) \partial_z + \frac{1}{2}Y'(z) + \frac{1}{2}Y'(z) u \partial_u \right) \psi_- \\
\delta_Y \psi_+ & = & \left(Y(z) \partial_z + Y'(z) + \frac{1}{2}Y'(z) u \partial_u \right) \psi_+
\ea
Similarly, for infinitesimal Lorentz transformations $\xi_{\Yb}$ parametrized by $\Yb(\zb) \partial_{\zb}$ one finds
\ba
\delta_{\Yb} \psi_- & = &   \left( \Yb(\zb) \partial_{\zb} + \Yb'(\zb) + \frac{1}{2}\Yb'(\zb) u \partial_u \right) \psi_-\\
\delta_{\Yb} \psi_+ & = &  \left( \Yb(\zb) \partial_{\zb} + \frac{1}{2}\Yb'(\zb) + \frac{1}{2}\Yb'(\zb) u \partial_u \right) \psi_+.
\ea
From a 2d perspective, the above expressions determine the holomorphic/antiholomorphic conformal dimensions of the ($u$-independent part of) $\psi_\pm$. In particular, one concludes the 2d spin (or equivalently the 4d  helicity) of  $\psi_{\pm}$ is equal to $\pm 1/2$.

The analogous calculation on the gauge fields leads to 
\ba
\delta_Y A_z & = & \left( Y(z) \partial_z + Y'(z) +   \frac{1}{2}Y'(z) u \partial_u \right) A_z  \\
\delta_Y A_{\zb} & =  &  \left( Y(z) \partial_z +   \frac{1}{2}Y'(z) u \partial_u \right) A_{\zb}  \\
\delta_{\Yb} A_z & = & \left( \Yb(\zb) \partial_{\zb} +    \frac{1}{2}\Yb'(\zb) u \partial_u \right) A_z \\
\delta_{\Yb} A_{\zb} & = & \left( \Yb(\zb) \partial_{\zb} +  \Yb'(\zb) + \frac{1}{2}\Yb'(\zb) u \partial_u \right) A_{\zb} \\
\ea
from which one concludes the 2d spin/4d helicity of  $A_z$ is $+1$ and that of $A_{\zb}$   is $-1$.

For holomorphic  vector fields, the generator  is given by 
\be
J_{Y} = \sum_{s=\pm} \int_{\I} du d^2 x  \left( \dot{A}_{-s} \delta_Y A_{s} + i \psib_s \delta_Y \psi_s \right) ,
\ee
with similar expression holding in the antiholomorphic case. These represent the  total angular momentum of the system evaluated at null infinity.

\subsubsection{Translations}
Spacetime translations (or more generally supertranslations) 
take the asymptotic form near null infinity
\be
\xi_f = f(z,\zb) \partial_u + \cdots
\ee
where  the sphere function $f$  for a translation $a^\mu$ is
\be
f(z,\zb) = a_\mu q^\mu(z,\zb) 
\ee
with $q^\mu$ given in \eqref{defq}. The action on the asymptotic fields is obtained from the Lie derivative, leading to
\be 
\delta_f \psi_s  = f \dot{\psi}_s, \quad \delta_f A_s  = f \dot{A}_s .
\ee
The phase space generator of the above transformation is then given by
\be
P_f = \sum_{s=\pm} \int_{\I} du d^2 x  \left( \dot{A}_{-s} \delta_f A_{s} + i \psib_s \delta_f \psi_s \right),
\ee
and represents the  total linear momentum of the system evaluated at null infinity.

\subsubsection{Large $U(1)$ gauge}
Gauge transformations \eqref{delLambda} with asymptotic behavior 
\be \label{fallLambda}
\Lambda(r,u,x) \stackrel{r\to \infty}{=} \lambda(x) + \cdots
\ee
induce the following action on the fields at null infinity:
\be \label{dellam}
\delta_\lambda A_a =  \partial_a \lambda, \quad \delta_\lambda \psi_{s} =- i e \lambda \psi_s.
\ee
The canonical generator for \eqref{dellam}  is given by
\be \label{Qlambda}
Q_\lambda  =  \int_{\I} du d^2 x \lambda \big(- \partial^a \dot{A}_a + e( \psib_+ \psi_+ + \psib_- \psi_-) \big).
\ee
By expressing the fermionic field in terms of Fock operators, the ``hard''  part of the charge can be written as 
\be
Q^\hard_\lambda = e \int d^2 x \lambda (\rho^\psi_+ + \rho^\psi_- - \rho^{\bar{\psi}}_+ - \rho^{\bar{\psi}}_-).
\ee
For $\lambda=1$, $Q_\lambda$  reduces to the total electric charge as measured at null infinity:
\be
Q_{\lambda=1}= e( N^\psi_+ + N^\psi_-) - e( N^{\bar{\psi}}_+ + N^{\bar{\psi}}_-),
\ee
where we recall that $N^X_\pm$ is the number operator defined in \eqref{defNX}.

\subsubsection{Axial rotations}
The axial symmetry \eqref{deltaA}  induces at null infinity the transformation
\be
\delta_\Ar \psi_s = -i s \psi_s.
\ee
The canonical generator is given by 
\be
Q_\Ar = ( N^\psi_+  +  N^{\bar{\psi}}_+ )  - (N^\psi_-  + N^{\bar{\psi}}_-)
\ee
and counts the excess of positive helicity fermions over negative helicity fermions.

\section{Soft electrons and fermionic asymptotic charges} \label{softelectron}
In this section we present a tree-level formula for soft electrons (and soft positrons)  and compute the associated asymptotic charges. For simplicity we treat all particles as outgoing, and correspondingly derive the asymptotic charge at  future null infinity. A detailed derivation of the soft electron theorem is given in appendix \ref{softparticlesapp}.

Consider a  tree-level amplitude involving $n$ hard particles and a  soft electron of  momentum $p^\mu = \w q^\mu(x)$, with $\w \to 0$ and $q(x)$ as in \eqref{defq}. As in other instances of  soft theorems, the dominant diagrams are those where the soft electron is attached to an external hard leg. There are two possibilities. Either the soft electron emerges from an external hard photon, leaving behind an internal electron line, or it emerges from an external hard positron, leaving behind an internal photon line:   
\begin{figure}[H]
\centering
\includegraphics[valign=m]{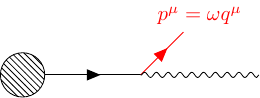} \quad \quad or \quad \quad \includegraphics[valign=m]{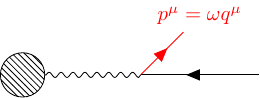} 
\caption{The two types of diagrams contributing to a soft electron amplitude.}
\end{figure}
In both cases one obtains a result that is proportional to the $n$-point amplitude left behind the soft emission vertex. However, unlike the situation for soft photons, 
there is a change in the type of hard particle involved in the process.\footnote{A change in the type of hard particles also occurs for soft  photinos \cite{stromfermionic} and for subleading soft photons in  presence of non-minimally coupled matter \cite{elvang2,mitra}.}  The proportionality factor in the two types of process  coincide, and is trivial unless there is certain helicity matching at the vertex. 
Taking for concreteness a soft electron of positive helicity and calling $p_i$ the momentum of the hard particle, the proportionality factor is given by (see appendix \ref{softparticlesapp})
\be \label{factor}
-e \frac{ \bar{u}_+(p) \slashed{\e}^+_i u_{+}(p_i)}{2 p \cdot p_i} = \frac{e}{\sqrt{\w \w_i}(z-z_i)}.
\ee
One thus arrives at the (positive helicty) soft electron theorem  
\be \label{softthmeplus}
\begin{aligned}
 \A_{n+1}(\{p_i \}, \w q^\psi_+))  \stackrel{\w \to 0}{=} & \frac{e}{\sqrt{\w}}  \sum_{i \in A_+} \frac{1}{\sqrt{\w_i}(z-z_i)}\A_n( \ldots,  {p_i}^A_{+} \to {p_i}^\psi_{+}, \dots) \\  +  & \frac{e}{\sqrt{\w}} \sum_{i \in \bar{\psi}_-} \frac{1}{\sqrt{\w_i}(z-z_i)}\A_n( \ldots,  {p_i}^{\bar{\psi}}_{-} \to {p_i}^A_{-} , \dots)  ,
\end{aligned}
\ee
where we use the labels $\psi, \bar \psi$ and $A$  for electrons, positrons and photons respectively and  particle helicities are displayed by $\pm$ subscripts. 
The argument in $\A_n$ indicates the  $n$-point amplitude involves a change in the  $i$-th hard particle type.

Following what has been done for other soft theorems, we can interpret the above formula as a Ward identity of  an asymptotic charge. Since the particle states in   \eqref{softthmeplus}  are normalized according to  the standard momentum-space Fock operators (see appendix  \ref{PBapp}), it is simpler to first write the charge in terms of them. From \eqref{softthmeplus} one can  read-off the direction-dependent fermionic charge
\begin{multline} \label{charge1stversion}
\frac{4 \pi i}{\sqrt{2}}Q_{\psi_+}(x) := \lim_{\w \to 0} \sqrt{\w} b^{standard}_+(\w q(x))  \\ - e \int \widetilde{d p'} \frac{1}{\sqrt{\w'}(z-z')} \left(  
a^{standard \, \dagger}_+(p') b^{standard}_+(p') + \  c^{standard \, \dagger}_-(p') a^{standard}_-(p')  \right).
\end{multline}
The overall normalization  is chosen for later convenience and the label \emph{standard} is to distinguish the  momentum-space Fock operators from the ones defined in section \ref{focksec}.

Next, we look to express  the charge  in terms of the asymptotic fields $\psi_s(u,x)$ and $A_s(u,x)$. To achieve this, we first rewrite \eqref{charge1stversion} in terms of the asymptotic Fock operators of section \ref{focksec}. These are related to the  momentum-space operators by (see appendix \ref{PBapp})
\be \label{relfockops0}
a^{standard}_s= 4 \pi i a_s , \quad    b^{standard}_s=  \frac{4 \pi i}{\sqrt{2 \w}} b_s , \quad c^{standard}_s=  \frac{4 \pi i}{\sqrt{2 \w}} c_s.
\ee
Substituting \eqref{relfockops0} in \eqref{charge1stversion} and using  $\dpp' = \frac{\w'}{2 (2 \pi)^3} d^2 x' d \w'$
we get
\begin{multline} \label{charge2ndversion}
Q_{\psi_+}(x) = \lim_{\w \to 0}  b_+(\w,x)  \\+ \frac{i e}{2 \pi} \int d^2 x' \frac{1}{(z-z')} \int_0^\infty \frac{d \w'}{2 \pi} \left(  
a^\dagger_+(\w',x') b_+(\w',x') + \  c^\dagger_-(\w',x') a_-(\w',x')  \right).
\end{multline}
We now Fourier transform from $\w$ to $u$-space using the expressions from section \ref{focksec}. Let us discuss the two terms in \eqref{charge2ndversion} separately. Following the standard terminology, we refer to them as ``soft'' and ``hard'' charge respectively. The ``soft'' part of the charge is found to be given by 
\be \label{Qsoftx}
Q^{soft}_{\psi_+}(x) \equiv \lim_{\w \to 0}  b_+(\w,x) = \int_{-\infty}^\infty du \psi_+(u,x) .
\ee
The hard term is a bit more involved. We start by rewriting it as
\be \label{Qhardx}
Q^{hard}_{\psi_+}(x) =i e \partial^{-1}_{\zb}  \sigma_{\zb +}(x),
\ee
where 
\be
\partial^{-1}_{\zb}  = \frac{1}{2\pi} \int d^2 x' \frac{1}{(z-z')} ,
\ee
and 
\be \label{defsigmazb}
\sigma_{\zb +}(x):=  \int_0^\infty \frac{d \w'}{2 \pi} \left(  
a^\dagger_+(\w',x') b_+(\w',x') + \  c^\dagger_-(\w',x') a_-(\w',x')  \right).
\ee
From Eqs. \eqref{fouriernull} and \eqref{fockops} one finds \eqref{defsigmazb} can be written as
\ba
\sigma_{\zb +}(x) & =&   \int_0^\infty \frac{d \w'}{2 \pi} \left(\tilde{A}_{\zb}(-\w',x') \tilde{\psi}_+(\w',x') + \tilde{\psi}_+(-\w',x')  \tilde{A}_{\zb}(\w',x') \right)  \\
 &=& \int_{-\infty}^\infty du A_{\zb}(u,x') \psi_+(u,x') . \label{sigmazb}
\ea
We finally combine the ``soft" and ``hard" charges by factoring out an inverse derivative  in the former
\be \label{Qsoftpsi}
Q^{soft}_{\psi_+}(x) = \partial^{-1}_{\zb} \int_{-\infty}^\infty du \partial_{\zb} \psi_+(u,x) .
\ee
Comparing \eqref{Qsoftpsi} with \eqref{Qhardx} and \eqref{sigmazb}, we conclude the total charge can be written as
\be \label{Qpsiplus}
Q_{\psi_+}(x)  =  \partial^{-1}_{\zb} \int^\infty_{-\infty} du  D_{\zb} \psi_+(u,x),
\ee
where
\be
D_a \psi_s \equiv (\partial_a + i e A_a) \psi_s
\ee
 is the gauge covariant derivative at null infinity.

Repeating the previous analysis for a negative helicity soft electron yields a charge of the form
\be \label{Qpsiminus}
Q_{\psi_-}(x)  =  \partial^{-1}_{z} \int^\infty_{-\infty} du  D_{z} \psi_-(u,x).
\ee
Finally, soft positrons lead to  charges that are the  complex conjugates of \eqref{Qpsiplus} and \eqref{Qpsiminus}. 

It is natural to combine all these  direction-dependent  charges  into a single smeared asymptotic charge which we define as 
\be \label{defFchi}
F_\chi := i  \int d^2 x du \left( \chib^{\zb}_+ D_{\zb} \psi_+  + \chib^{z}_- D_z \psi_-  + \chi^z_+ D_z \psib_+ + \chi^{\zb}_- D_{\zb} \psib_-  \right) .
\ee
where $\chi^z_+(x)$ and $\chi^{\zb}_-(x)$ are the components of a spinor-vector smearing parameter
\be
\chi= (\chi^z_+,\chi^{\zb}_-), 
\ee
with complex conjugate  $\chib=(\chib^{\zb}_+,\chib^{z}_-)$. 
We take $\chi$ to be grassmanian  so that $F_\chi$ is real/hermitian.

\section{On fermionic asymptotic symmetries} \label{fermionicsymsec}
Even though conserved asymptotic charges imply the existence of asymptotic symmetries, the nature of the latter may  be challenging to decipher (see for example  \cite{stromLow,stromfermionic}). One reason for this difficulty is that asymptotic symmetries are  spontaneously  broken \cite{stromlectures}, while the charges  obtained from  soft theorems are evaluated on 
a single vacuum sector. 
To characterize the symmetry  one needs to make manifest the vacuum manifold,  thus going beyond standard  Fock-space amplitudes  \cite{Kapec:2017tkm,Choi:2017ylo,Prabhu:2022zcr,Kapec:2022hih,allthat}.  At the classical level, this usually requires an extension of the radiative phase space, see e.g. \cite{Kapec:2014zla,Freidel:2019ohg,perazanagy,Peraza:2023ivy,Sudhakar:2023uan,Aggarwal:2018ilg,Campiglia:2020qvc,Capone:2023roc,Chowdhury:2022gib}.

In this section we take  the first steps towards characterizing the asymptotic fermionic symmetry implied by the soft electron theorem. We will start by evaluating the charge action according to the radiative phase space brackets \eqref{elemPBs}. We shall see this action does not respect the  $|u| \to \infty$ behavior of the asymptotic fields, thus indicating the need of a phase space extension.   
We leave for future work the identification of such extension as well as the related problem of fully characterizing the symmetry algebra. 
We will nevertheless be able to verify the commutator  between fermionic and bosonic symmetries. In appendix \ref{subphapp} we shall further discuss a naive evaluation of the commutator between two fermionic symmetries.

\subsection{Symmetry action from radiative PBs}
Infinite dimensional phase spaces may present  subtleties that are absent in finite dimensions. 
In particular, not all phase-space functionals are guaranteed to yield well-defined PB actions. As we shall see, this is the case for the fermionic charge $F_\chi$. 

Let us for a moment ignore the aforementioned subtlety and consider the action obtained from the standard formula
\be \label{defdeltachiPB}
\delta_\chi = \{ \cdot, F_\chi \}.
\ee
From the elementary PBs \eqref{elemPBs} and the expression for $F_\chi$  \eqref{defFchi} one finds\footnote{We recall that $D_a = \partial_a + i e A_a$ is the gauge covariant derivative at null infinity and  $\chi_{z -} = \chi^{\zb}_-, \quad \chi_{\zb +} = \chi^{z}_+$. The action on $\psib_\pm$ can be obtained from that of $\psi_\pm$ by  complex conjugation. To simplify expressions  we have chosen to display $ \delta_\chi \dot A_s$ rather than $\delta_\chi A_s$ (which is non-local in $u$).}
\be \label{delchiradPB}
\begin{aligned}
\delta_\chi \psi_+ &=  D_z \chi^z_+\\
 \delta_\chi \psi_- &= D_{\zb} \chi^{\zb}_- \\
 \delta_\chi \dot A_z &=  \frac{e}{2}(  \chib_{z +} \psi_+ - \chi_{z -} \psib_- )\\
  \delta_\chi \dot A_{\zb} &=  \frac{e}{2}(\chib_{\zb -} \psi_- - \chi_{\zb +} \psib_+ ).
\end{aligned}
\ee
We first notice that $\lim_{u \to \pm} \delta_\chi \psi_s \neq 0$, and thus the transformation does not preserve the condition $\psi_s \stackrel{|u| \to \infty}{\to} 0$ \eqref{fallupsi}. This suggests   a phase space extension that allows for non-trivial asymptotic values of $\psi_s$ when $u \to \pm \infty$. 

The transformation rule for  $\dot A_s$ \eqref{delchiradPB} is compatible with the asymptotic behaviour of $A_s$ given in \eqref{falluA} provided $\psi_s$ decays to zero as in \eqref{fallupsi}.  The discussion from the previous paragraph however indicates that in an extended space where   $\lim_{u \to \pm \infty}  \psi_s \neq 0$ we would need to allow for non-trivial $|u| \to \infty$ values of $\dot A_s$. This resembles the situation for the asymptotic charges obtained from subleading photons \cite{stromLow}, whose action creates an $O(u)$ term in $A_s$. 

If one continues the previous considerations back and forth between  $\delta_\chi \psi_s$ and $\delta_\chi A_s$,  one is led to  conclude that all powers of $u$ should be allowed in 
  the $|u| \to \infty$ behaviour of the fields.  We recall however that  \eqref{delchiradPB} is at best valid only around the trivial vacuum and cannot be trusted beyond linear order in $\chi$.  Higher order iterations of $\delta_\chi$ 
 may require the inclusion of terms that are absent in \eqref{delchiradPB} and that could modify the previous conclusion. 
We leave for future work the elucidation of this non-linear structure.\footnote{A preliminary attempt go beyond linear order is presented in appendix \ref{subphapp}, where we evaluate the commutator  between two fermionic variations \eqref{delchiradPB}. Although the result cannot be trusted due to the aforementioned limitations, it may still be of use in more complete  treatments.}
There are  however  checks to be made at first order in $\chi$, namely the commutation of $\delta_\chi$ with the bosonic symmetries reviewed in section \ref{bossymsec}. We discuss them next.

\subsection{Algebra relations with bosonic symmetries} \label{consistencysec}
Let $\delta_B$ be a bosonic symmetry action corresponding to a charge $B$, i.e.
\be \label{deltaBPB}
\delta_B = \{ \cdot, B \}.
\ee
We restrict attention to the bosonic symmetries discussed in section \ref{bossymsec} so that  $\delta_B = \delta_Y, \delta_f, \delta_\lambda, \delta_{\Ar}$ for $B= J_Y, P_f, Q_\lambda, Q_\Ar$ respectively. As we shall see, these symmetries have a natural action on the fermionic parameter,
\be
\chi \mapsto \chi +\delta_B \chi
\ee
such that the commutator of variations is given by
\be \label{deltaalg}
[\delta_\chi , \delta_B ] = \delta_{\delta_B \chi}.
\ee
At the level of charges, this should imply the PB relations\footnote{In the absence of a central extension, as it  turns out to be the case.} 
\be \label{PBFB}
\{F_\chi, B \}  = - F_{\delta_B \chi} . 
\ee
There are however various subtleties in the evaluation of PBs that make \eqref{PBFB} challenging  to verify directly.  It is for this reason that we will instead focus on the 
relations
\ba
\delta_B F_\chi & = &  - F_{\delta_B \chi} ,\label{delBF} \\
\delta_\chi B & = &  F_{\delta_B \chi} \label{delchiB}.
\ea
It turns out  that \eqref{delBF}  follows straightforwardly from the expression of $F_\chi$ and \eqref{deltaalg}. This should lead, via  \eqref{deltaBPB},  to \eqref{PBFB}. Finally from \eqref{defdeltachiPB} one would arrive at \eqref{delchiB}. There are however two obstructions to this logic chain.

First, Eq. \eqref{deltaBPB} does not hold for $Q_\lambda$ if one uses the radiative PBs \eqref{elemPBs} (see the discussion following this equation). Since these were the brackets used in the definition of $\delta_\chi$   \eqref{delchiradPB},
there will be a mismatch in \eqref{delchiB} when $B=Q_\lambda$. This problem should go away if $\delta_\chi$ is constructed from improved PBs as the ones proposed in \cite{stromQED}.

A second obstruction appears from the fact that the relation
\be
\delta F_\chi = \Omega(\delta,\delta_\chi)
\ee
 only holds for  variations $\delta$ that decay to zero faster than the assumed  ones in order  to compensate for the singular behavior of $\delta_\chi$ at $u = \pm \infty$. This leads to additional difficulties in verifying \eqref{delchiB}, for instance when $B=J_Y$. A complete fix to this second obstruction would presumably require the inclusion of soft fermionic degrees of freedom in the symplectic structure.

We now discuss in more detail the situation for each bosonic symmetry separately.

 \subsubsection{Lorentz}
The transformation properties of $\chi$ under the Lorentz group follow directly from those of the elementary fields $A_s$ and $\psi_\pm$. By direct evaluation one readily obtains \eqref{deltaalg} and \eqref{delBF} with
\ba
\delta_{(Y,\Yb)} \chi^z_{+} & = \left(Y \partial_z + \Yb \partial_{\zb} + \frac{1}{2}\Yb' \right) \chi^z_{+}&  \\
\delta_{(Y,\Yb)} \chi^{\zb}_{-}& = \left(Y \partial_z + \Yb \partial_{\zb} + \frac{1}{2}Y' \right)  \chi^{\zb}_{-}& .
\ea
Thus, from a 2d perspective,  $\chi^z_{+}$  and  $\chi^{\zb}_{-}$ have (anti-)holomorphic dimensions  $(h,\hb)=(0,1/2)$ and $(1/2,0)$ respectively.

Relation \eqref{delchiB} is verified provided one discards boundary terms of the type 
\be
\left[A_{\zb} \delta_\chi \delta_{(Y,\Yb)} A_z \right]^{u=\infty}_{u=-\infty},  
\ee
which are however non-trivial under the assumed  fall-offs \eqref{falluA}, \eqref{fallupsi}. 

 \subsubsection{Translations}
It is easy to verify that  $\delta_\chi$ commutes with translations,
\be
[\delta_\chi,\delta_f]=0
\ee
and that $\delta_f F_\chi=0$ (assuming the fall-offs \eqref{falluA}, \eqref{fallupsi} otherwise there could be non-zero  boundary terms). The relation $\delta_\chi P_f=0$ follows with no caveats.
\subsubsection{Large $U(1)$ gauge}
As in  the Lorentz case, the transformation properties of $\chi$ follow from those of the elementary fields. One can easily verify \eqref{deltaalg} and \eqref{delBF} with
\ba
\delta_\lambda \chi^z_{+}  & = & - i e  \chi^z_{+} \\
\delta_\lambda \chi^{\zb}_{-}  & = & - i e  \chi^{\zb}_{-}.
\ea
As anticipated earlier, Eq. \eqref{delchiB} is not verified unless one addresses the $1/2$ discontinuity in the radiative PBs \eqref{discontinuity}. Let us describe how the issue arises in the present context. 

The RHS of \eqref{delchiB} takes the form 
\ba
F_{\delta_\lambda \chi} &= &  -e \int_{\I} \lambda \chib^{\zb}_+ D_{\zb} \psi_+ + \cdots  \\
&= & - e \int_{\I} \partial_{\zb} \lambda \chib^{\zb}_+  \psi_+ + e \int_{\I}  \lambda D_{\zb} \chib^{\zb}_+  \psi_+ + \cdots, \label{Fdellamchi2}
\ea
where in the second line we integrated by parts and for simplicity we are only displaying  what corresponds to the first   term  in $F_\chi$ \eqref{defFchi} (the others follow a similar pattern).

The LHS of \eqref{delchiB} can be written as a
\be
\delta_\chi Q_\lambda = \delta_\chi Q^\hard_\lambda + \delta_\chi Q^\soft_\lambda
\ee
with
\ba
\delta_\chi Q^\hard_\lambda & = & \delta_\chi  \int_{\I}  e \lambda(\psib_+ \psi_+ + \psib_- \psi_-) \\
& = &  e \int_{\I}  \lambda \delta_\chi \psib_+ \psi_+ + \cdots \\
& = &  e \int_{\I}  \lambda D_{\zb} \chib^{\zb}_+  \psi_+ + \cdots 
\ea
and
\ba
\delta_\chi Q^\soft_\lambda & = &   \int_{\I} (\partial_{\zb} \lambda \dot{A}_{z} + \partial_z\lambda  \dot{A}_{\zb} )\\
& = &  \int_{\I} \partial_{\zb} \lambda \delta_\chi  \dot{A}_{z}  + \cdots \\
& = &  \frac{e}{2}\int_{\I} \partial_{\zb} \lambda  \chib^{\zb}_{+} \psi_+ + \cdots,
\ea
where we  only displayed the terms that correspond to those shown in \eqref{Fdellamchi2}. Comparing the expressions one sees that whereas $\delta_\chi Q^\hard_\lambda$ reproduces the first term in \eqref{Fdellamchi2}, $\delta_\chi Q^\soft_\lambda$ fails to reproduce the second term by a factor of $1/2$. The origin of this mismatch is the same as the one leading to the discontinuity discussed in Eq. \eqref{discontinuity} and so it  would be fixed by the use of improved PBs \cite{stromQED}.

\subsubsection{Axial rotations}
In this case, equations \eqref{delBF} and \eqref{delchiB} are verified with no caveats, with
\ba
\delta_\Ar \chi^z_{+}  & = & - i   \chi^z_{+} \\
\delta_\Ar \chi^{\zb}_{-}  & = &  i   \chi^{\zb}_{-}.
\ea

\section{Outlook} \label{outlook}
Over the past decade there has been a fruitful revision on the subject of asymptotic symmetries in (asymptotically) flat spacetimes, driven by the discovery of their  connection with soft theorems and memory effects \cite{stromlectures}. Whereas  asymptotic symmetries were originally tied to gauge symmetries that are non-trivial at infinity, their relation to soft theorems led to a broader perspective. Indeed, there is a growing list of soft theorems that admit an interpretation in terms of asymptotic symmetries with no obvious  gauge origin.\footnote{What appears as a non-gauge asymptotic symmetry may sometimes be realized as large gauge, either by allowing divergent gauge transformations \cite{subphotonAL} or by performing a change of field variables \cite{ronak}. It is however not clear whether such  reinterpretation is always possible.}   Here we have enlarged such list, by presenting an asymptotic symmetry associated to  soft electrons in tree-level massless QED.\footnote{Although we have set our discussion in the context of  QED, it should admit a straightforward generalization to the case of non-abelian gauge fields as well as chiral fermions.} Following what is  by now a standard procedure, we identified the form of the asymptotic charge and initiated the study of its symmetry action. 

To our knowledge, this is the first example of a fermionic asymptotic symmetry in a theory with no \emph{standard} supersymmetry. The existence of such fermionic symmetry, however,  suggest  that tree-level massless QED possess an \emph{asymptotic} supersymmetry algebra. Unfortunately the present work remains agnostic as to what such algebra should be. Below we describe this and others shortcomings of our analysis as well as possible strategies to overcome them. 

We followed what may be referred to as a canonical approach to asymptotic symmetries, in which the classical phase space at null infinity provides the bridge between symmetries and charges. 
A difficulty with this approach is that it often requires an enlarged version of the radiative phase space, with no simple recipe to determine it.\footnote{Recent developments on so called corner symmetries \cite{donnellyfreidel,geiller,Freidel:2021dxw,leigh} may in fact yield such recipe (at least in the case of large gauge symmetries).} Whereas we showed that an enlargement is implied by soft electrons, we were not able to characterize it beyond linear level. This in turn impeded us to trustfully evaluate  the commutator between fermionic asymptotic symmetries.

 
There is however  a second approach in which symmetries are described in terms of   2d conserved currents of a dual  CFT \cite{Barnich:2013axa,stromYM,sundrum}. There, the algebraic structure is read off from collinear factorization theorems \cite{TaylorOPE,Taylor2,stromOPE,holosymalg,OPEallspins,jiang,shamik1,shamik2} without the need of phase-space considerations.\footnote{The price to pay, however, is that so far such approach only captures  symmetry relations among generators of the same helicity sign. See \cite{freidelhs,freidelYM,Donnay:2018neh,puhm,Crawley:2021ivb} for discussions  on how  the canonical and CFT approaches relate to each other.} It would be very interesting to study the  fermionic symmetries presented here from this perspective, as it could allow us to extract information on the  fermionic symmetry algebra.

Regardless of the approach,  it has been found  in the context of gauge and gravitational theories  that completeness of the asymptotic symmetry algebra requires  charges associated to subleading soft theorems of arbitrarily high order, see e.g. \cite{clspatial,freidelhs}. It would be interesting to explore the existence of higher order soft electron theorems and their connection with the soft photon ones. A hint that a nontrivial interplay may occur is provided by a naive evaluation of the commutator between the fermionic variations  presented here, leading to 
an expression that is reminiscent to the subleading soft photon charge action (see appendix \ref{subphapp}).

A final pressing open problem regards the fate of the symmetries beyond tree-level. The structure of their modification or breaking due to loop corrections should be worth studying. In particular, it would be interesting to explore any possible connection with the chiral anomaly, whose consequences at null infinity were recently analyzed  in \cite{agullo}.

\section*{Acknowledgements}
We would like to thank Iván Agulló, Alok Laddha, Guzmán Hernández, Pablo Pais and Michael Reisenberger for illuminating discussions. We specially thank Alok Laddha for his feedback and encouragement. AA acknowledges support from PEDECIBA and from CSIC grant I+D 583.  MC acknowledges support from PEDECIBA and from ANII grant FCE-1-2019-1-155865.  This research was supported in part by Perimeter Institute for Theoretical Physics. Research at Perimeter Institute is supported by the Government of Canada through the Department of Innovation, Science and Economic Development Canada and by the Province of Ontario through the Ministry of Research, Innovation and Science.

\appendix

\section{Asymptotic vs. momentum-space Fock operators} \label{PBapp}
In this appendix we work out the relation  between the Fock operators introduced in section \ref{focksec} and the standard momentum-space Fock operators. In particular we verify they yield equivalent (anti) commutation relations.

We start with the momentum expansion of asymptotic free fields:
\be \label{momentumexpansion}
\begin{aligned}
\A_\mu(X) & =  \sum_{s=\pm} \int \dpp \left( a_s(p) \e^{s *}_{\mu}(p) e^{i p \cdot X}  + a^\dagger_s(p) \e^{s}_{\mu}(p) e^{-i p \cdot X} \right) \\
\Psi(X) &=  \sum_{s=\pm} \int \dpp \left( b_s(p) u_s(p) e^{i p \cdot X} + c^\dagger_s(p) v_s(p) e^{-i p \cdot X} \right)
\end{aligned}
\ee
where $\dpp = d^3 p /((2 \pi)^3 2 |p|)$  and the photon/electron polarization vectors are described below.  To simplify notation we do not yet include additional labels to the Fock operators, but it should be kept in mind that they have different normalization from those introduced in section \ref{focksec}.

To study the null infinity limit of these expressions, we parametrize $X^\mu$ in terms of  $(r,u,x)$ as in \eqref{Xitorux} and write the null momentum $p^\mu$ as
\be
p^\mu = \w q^\mu(x')
\ee
with $q^\mu$ as in \eqref{defq} with $x'=(z',\zb')$. In this parametrization the momentum measure takes the form
\be
\dpp = \frac{\w}{2 (2 \pi)^3} d^2 x' d \w
\ee 
and  the  plane-wave phase becomes
\be
p \cdot X  = r \w q(x) \cdot q(x') + u \w k \cdot q(x')
\ee
where
\be
 q(x) \cdot q(x') = -|z-z'|^2.
\ee
In the $r \to \infty$ limit, the $d^2 x'$ integral can be evaluated by a saddle point analysis. In the simplest case of a  positive-frequency scalar  one gets
\be
\lim_{r \to \infty} \int \dpp \phi(p)  e^{i p \cdot X} =   \frac{ 1}{4 \pi i r} \int^{\infty}_0 \frac{d \w}{2 \pi} \phi(\w q(x)) e^{- i \w u} + O(1/r^2).
\ee
Before extending this formula to the photon and fermion fields, let us recall the form of the corresponding momentum wave-functions. The polarization vector for a photon with momentum $p=\w q^\mu(x)$ can be taken to be \cite{stromQED}
\be \label{photoneps}
 \e^{+ \mu} = \partial_{z} q^\mu = \frac{1}{\sqrt{2}}(\zb,1,-i,-\zb) , \quad \e^{-\mu} = \partial_{\zb} q^\mu = \frac{1}{\sqrt{2}}(z,1,i,-z) 
\ee
which satisfy
\be \label{epsids}
 \e^\pm  \cdot q =0, \quad \e^{\pm} \cdot \e^{\pm}=0, \quad \e^+ \cdot \e^{-} =1, \quad \e^+_\mu \e^-_\nu +\e^-_\mu \e^+_\nu + \frac{q_\mu k_\nu+k_\mu q_\nu}{q \cdot k} = \eta_{\mu \nu}
\ee
with $k^\mu$ a null vector with non-zero dot product with $q^\mu$, as for instance the one given in  \eqref{defk}. For the fermion we have (see e.g. \cite{elvang,stromfermionic})
\be \label{spinorsuv}
u_+ = v_- = 2^{1/4} \sqrt{\w} \begin{pmatrix} 0 \\ 0 \\ 1 \\ z   \end{pmatrix} , \quad  u_- = v_+ = 2^{1/4} \sqrt{\w}\begin{pmatrix} -\zb  \\ 1 \\ 0 \\ 0  \end{pmatrix} 
\ee
which satisfy
\be \label{idsus}
\slashed{q} u_{\pm} =0, \quad \ub_s \gamma^\mu u_{s'} = 2 p^\mu \delta_{s s'}, \quad \sum_{s} u_s \bar{u}_s = - \slashed{p}, 
\ee
and similarly for $v_{\pm}$. Expressions \eqref{photoneps}, \eqref{spinorsuv} can be  obtained from those used in the spinor-helicity formalism \cite{elvang} with the choice of spinor $2^{1/4}\sqrt{\w} (1,z)$  for the momentum  $p^\mu = \w q^\mu(z,\zb)$.\footnote{The photon polarization in \eqref{photoneps} corresponds to a choice of  ``reference spinor'' of the form $(1,\infty)$.}

Using the above expressions, one finds
\be
\A_z =\frac{1}{4 \pi i}  \int^{\infty}_0 \frac{d \w}{2 \pi}\left( a_+ e^{- i \w u} - a^\dagger_- e^{i \w u} +  \right) + O(1/r)
\ee
\be
\Psi = \frac{1}{4 \pi i r}  \int^{\infty}_0 \frac{d \w}{2 \pi} 2^{1/4} \sqrt{\w}   \begin{pmatrix} \begin{array}{c} -\zb \\ 1 \end{array}(  b_- e^{-i \w u}- c^\dagger_+ e^{i \w u} ) \\ \begin{array}{c}  1 \\ z \end{array} (b_+e^{-i \w u}- c^\dagger_- e^{i \w u} )\end{pmatrix} .
\ee
Comparing with \eqref{fouriernull}, \eqref{fockops} we see that the  momentum space Fock operators are related to the Fock operators of section \ref{focksec} by
\be \label{relfockops}
a_s =  \frac{1}{4 \pi i}  a^{standard}_s, \quad  b_s =  \frac{\sqrt{2 \w}}{4 \pi i} b^{standard}_s, \quad c_s =  \frac{\sqrt{2 \w}}{4 \pi i} c^{standard}_s,
\ee
where we have added the label ``standard"  to the operators of this section to distinguish them from those of section \ref{focksec}. The commutation relations \eqref{elemcomms} then imply the standard momentum-space (anti) commutators
\be 
[d^{standard}_s(p), d^{standard \, \dagger}_{s'}(p')] = 2  |p| (2 \pi)^3 \delta_{s s'} \delta^{(3)}(p-p'), \quad d=\{a,b,c\},
\ee
where we used that  $ \delta^{(3)}(p-p') = \frac{1}{\w^2} \delta(\w - \w') \delta^{(2)}(x,x')$ for $p= \w q(x)$ and $p'= \w q(x')$.

\section{Soft photons and soft electrons} \label{softparticlesapp}
In this appendix we derive the tree-level formulas for the emission of  soft particles.  We treat all particles as outgoing. 

Since the  formulas are dictated by the  QED 3-point interaction,  it will  be useful to start  the discussion by recalling the structure of general  3-point amplitudes. Consider then a process involving an electron, positron and photon of momenta 
\be
\begin{aligned}
p_e & = \w_e q(z_e,\zb_e) \\
p_{\bar e} & = \w_{\bar e} q(z_{\bar e},\zb_{\bar e})\\
p_{\gamma} & = \w_\gamma q(z_{\gamma},\zb_{\gamma}) 
\end{aligned}
\ee
 and helicities $s, \bar s$ and $h$. 
 The (momentum conserving delta function stripped) amplitude is given by
\be
A_3((p_e,s),(p_{\bar e},\bar s), (p_\gamma,h)) =
 \includegraphics[valign=m]{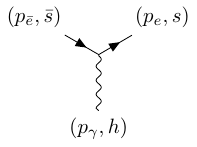}  = - i e \bar{u}_s(p_e) \gamma^\mu v_{\bar{s}}(p_{\bar e}) \e^h_{\mu}(p_\gamma).
\ee
In order for the amplitude to be non-zero we must have $\bar{s}=-s$ whereas  $h$ can be arbitrary. Writing the 4 posibilites in holomorphic coordinates (using the expressions for polarization vectors given in the previous section) one finds
\be
\begin{aligned} \label{A3}
A_3((p_e,+),(p_{\bar e},-), (p_\gamma,+)) &=  2 i e \sqrt{\w_e \w_{\bar e}} (\bar{z}_\gamma-\bar{z}_{e})  \\
A_3((p_e,+),(p_{\bar e},-), (p_\gamma,-)) &=  2 i e \sqrt{\w_e \w_{\bar e}} (z_\gamma-z_{\bar e}) \\
A_3((p_e,-),(p_{\bar e},+), (p_\gamma,+)) &=  2 i e \sqrt{\w_e \w_{\bar e}} (\bar{z}_\gamma-\bar{z}_{\bar e})\\
A_3((p_e,-),(p_{\bar e},+), (p_\gamma,-)) &=  2 i e \sqrt{\w_e \w_{\bar e}} (z_\gamma-z_{e}) .
\end{aligned}
\ee
We have not yet imposed momentum conservation. This typically leads to trivial amplitudes (unless one allows for complex momenta \cite{elvang}). An exception however is when one of the particles goes soft with the remaining two becoming coincident:\footnote{Since we  are working with real momenta,  $z \to z_0$ implies $\zb \to \zb_0$.}
\be \label{omegastozero}
\begin{aligned}
\w_\gamma \to 0 & \implies  z_{\bar e} \to z_e , \quad \w_{\bar e} \to \w_e \\
\w_e \to 0  & \implies z_{\bar e} \to z_\gamma , \quad \w_{\bar e} \to \w_\gamma \\
\w_{\bar e} \to 0  & \implies z_{ e} \to z_\gamma , \quad \w_{e} \to \w_\gamma . 
\end{aligned}
\ee
Notice that in the soft electron/positron limit, two out of the four amplitudes  in \eqref{A3} vanish.

We next describe the behavior of general amplitudes when one particle goes soft. As is well known, the leading contribution comes from diagrams where the soft particle is attached to an external hard particle. In the following we  discuss the relevant diagrams for each type of soft particle. We consider processes involving $n$ hard particles and $1$ soft particle.


\subsection{Soft photon}
The amplitude for a process in which a  photon is attached to an external electron is given by
\be \label{softphotonfromelectron}
\begin{aligned}
\includegraphics{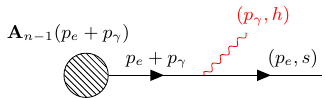}     & = - i e \e_\mu^h(p_\gamma) \ub_s(p_e) \gamma^\mu i \frac{\slashed{p}_e+\slashed{p}_\gamma}{(p_e+p_\gamma)^2} \Arb_{n-1}(p_e+p_\gamma)  \\
& \stackrel{\w_\gamma \to 0}{=}  e \e_\mu^h(p_\gamma) \ub_s(p_e) \gamma^\mu  \frac{\slashed{p}_e}{2 p_e \cdot p_\gamma} \Arb_{n-1}(p_e) +O(\w_\gamma^0) \\
& =  - e \frac{\e^h(p_\gamma)\cdot p_e}{p_\gamma \cdot p_e} A_n + O(\w_\gamma^0)
\end{aligned}
\ee
where $\Arb_{n-1}$ denotes the spinor-valued off-shell amplitude corresponding to the  remaining $n-1$ hard particles. In going to the second line we kept the leading terms in the $\w_\gamma \to 0$ limit, whereas in going to the third line we used the identities in \eqref{idsus} and the fact that the fermion helicities in the vertex must match for the amplitude to be nonzero. $A_n$ represents the amplitude for the  $n$ hard particles.

One can similarly compute the amplitude for the case where the photon is attached to an outgoing positron, resulting in an expression as the above  modulo an overall sign. Summing over all external particles one obtains Weinberg's soft photon theorem \cite{weinberg}.

\subsection{Soft electron}
We now consider processes in which an electron goes soft. From the structure of the QED vertex, we see that the electron can be attached either to an external photon or to an external positron (recall we are taking all particles to be outgoing). 

The contribution when the  electron is attached to an external photon is
\be
\begin{aligned}
\includegraphics{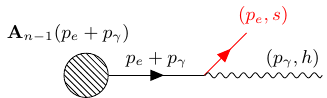}   &= - i e \e_\mu^h(p_\gamma) \ub_s(p_e) \gamma^\mu i \frac{\slashed{p}_e+\slashed{p}_\gamma}{(p_e+p_\gamma)^2} \Arb_{n-1}(p_e+p_\gamma) \\
& \stackrel{\w_e \to 0}{=}  e \e_\mu^h(p_\gamma) \ub_s(p_e) \gamma^\mu  \frac{\slashed{p}_\gamma}{2 p_e \cdot p_\gamma} \Arb_{n-1}(p_\gamma) +O(\w_e^0) \\
& =  - e \frac{\e_\mu^h(p_\gamma) \ub_s(p_e) \gamma^\mu u_s(p_\gamma)}{2 p_e \cdot p_\gamma} A_{n}((\gamma, p_\gamma,h) \to (e, p_\gamma,s)) +O(\w_e^0) .
\end{aligned}
\ee
Notice that the starting point is the same as in the soft photon case discussed in \eqref{softphotonfromelectron}. The difference is that  we are now taking the electron, rather than the photon, to be soft.  Again one obtains
an expression 
that is proportional to an $n$-point amplitude $A_n$.
However, unlike the soft photon theorem,  the external hard particle involved in the process  is replaced by a different type of particle (in this case a photon is replaced by an electron, as  indicated in the argument of $A_n$).   The multiplicative factor in front of  $A_n$ can be  simplified by writing the momenta and polarization vectors in holomorphic coordinates. The only non-trivial case is when $h$ and $s$ are of the same sign (this can be seen from Eqs. \eqref{A3} and \eqref{omegastozero}) and one finds
\ba
  - e \frac{\e_\mu^+(p_\gamma) \ub_+(p_e) \gamma^\mu u_+(p_\gamma)}{2 p_e \cdot p_\gamma}  &=&  \frac{e}{\sqrt{\w_e \w_\gamma}(z_e-z_\gamma)},\\
  - e \frac{\e_\mu^-(p_\gamma) \ub_-(p_e) \gamma^\mu u_-(p_\gamma)}{2 p_e \cdot p_\gamma} &=& \frac{e}{\sqrt{\w_e \w_\gamma}(\zb_e-\zb_\gamma)}.
\ea

The contribution when the electron is attached to an external positron is 
\be
\begin{aligned}
\includegraphics{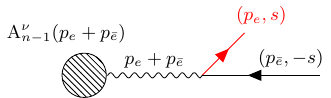}  &= - i e \ub_s(p_e) \gamma^\mu v_{-s}(p_{\bar e})  \frac{-i \eta_{\mu \nu}}{(p_e+p_{\eb})^2} \Ar^\nu_{n-1}(p_e+p_{\eb}) \\
& \stackrel{\w_e \to 0}{=} - e \ub_s(p_e) \gamma^\mu u_{s}(p_{\bar e})  \frac{\sum_h \e^h_{\mu}(p_{\eb}) \e^{-h}_{\nu}(p_{\eb})}{2 p_e\cdot p_{\eb}} \Ar^\nu_{n-1}(p_{\eb}) +O(\w_e^0) \\
& =  - e \frac{\e_\mu^s(p_{\eb}) \ub_s(p_e) \gamma^\mu u_s(p_{\eb})}{2 p_e \cdot p_{\eb}} A_{n}((\eb, p_{\eb},-s) \to (\gamma, p_{\eb},-s)) +O(\w_e^0) .
\end{aligned}
\ee
Here we assumed the positron helicity is opposite to that of the soft electron, since otherwise the amplitude vanishes.  $ \Ar^\nu_{n-1}$ is a vector-valued off-shell amplitude associated to the $n-1$ hard particles not involved in the process under consideration. In going to the second line we used the  resolution of the identity \eqref{epsids} together with $\slashed{p} u(p) =0$ and $p_\nu \Ar^\nu_{n-1}(p)=0$. Finally, in the third line, we used the fact that only the $h=s$ photon polarization yields a non-zero contribution.

We can finally combine both type of contributions to obtain the ``soft electron theorem" of Eq. \eqref{softthmeplus}. 


\section{Naive commutator of fermionic charges and subleading soft photons} \label{subphapp}
In this appendix we present the asymptotic symmetry associated to Low's subleading photon theorem \cite{Low}, as first introduced by Lysov, Pasterski and Strominger (LPS) \cite{stromLow}. We next evaluate the commutator of two fermionic symmetries and compare it with the  LPS symmetry. Even though the two expression do not coincide, there are certain similarities that suggest the LPS symmetry may feature in an eventual (super) symmetry algebra involving the fermionic generators. We hope the calculation presented here will be of use in a more complete treatment.

\subsection{Asymptotic symmetries for subleading soft photon} \label{LPSapp}
Soft photons obey a factorization theorem at subleading order \cite{Low} that can be  understood  in terms of  LPS  asymptotic symmetries \cite{stromLow}. The corresponding charges are parametrized by vector fields $Y^a$ on the sphere and take the form,
\be
Q^{\lps}_Y = \int_\I du d^2 x \left( -2 u \dot{A}_{\zb} \partial_z^2 Y^z + u \partial_z Y^z j_u + Y^z j_z \right) + c.c.,
\ee
where $j_u$ and $j_a$ are the leading components of the asymptotic current,  
\ba
\J_u &= & j_u/r^2 + \cdots \label{jufall} \\
\J_a & = & j_a/r^2 +\cdots. \label{jafall}
\ea
In order to compute the symmetry action on the spinor field, we need to express $j_u$ and $j_a$ in terms of $\psi_s$. This is achieved by substituting the asymptotic expansion of the Dirac spinor \eqref{asymPsi} in the current expression \eqref{defJ}. For the $u$ component one gets
\be
j_u = -e \sum_s \psib_s \psi_s.
\ee
The evaluation of $j_a$ requires  one order further  in the asymptotic expansion of the Dirac field,
\be \label{asymPsi2}
\Psi = \frac{1}{r} \ov{0}{\Psi}+ \frac{1}{r^2} \ov{1}{\Psi} + \cdots 
\ee
where  $\ov{0}{\Psi}$ is the leading term displayed in \eqref{asymPsi}. 
Substituting \eqref{asymPsi2} in the current expression one finds
\ba \label{Ja}
\J_a & =& \frac{e}{r} \left( \ov{0}{\Psib}+ \frac{1}{r} \ov{1}{\Psib} + \cdots \right)  \partial_a \slashed{q} \left( \ov{0}{\Psi}+ \frac{1}{r} \ov{1}{\Psi} + \cdots \right)  \\
& =& \frac{e}{r^2}  \left( \ov{0}{\Psib}  \partial_a \slashed{q} \ov{1}{\Psi}  + \ov{1}{\Psib}  \partial_a \slashed{q} \ov{0}{\Psi}\right) + \cdots
\ea
where we used that 
\be \label{cond1}
 \ov{0}{\Psib}  \partial_a \slashed{q} \ov{0}{\Psi}  =0.
\ee
We finally need to express $\ov{1}{\Psi}$ in terms of  $\psi_s$. Imposing the asymptotic free-field equation on $\Psi$ one finds\footnote{Whereas this suffices for tree-level considerations, a more complete treatment should deal with the full non-linear asymptotic field equations. This in turn may modify the assumed fall-offs \eqref{asymPsi2} (and hence \eqref{jufall}, \eqref{jafall}). See \cite{sayali} for related discussions.}  
\be \label{Diraceqinr}
0  =  \slashed{\partial} \Psi = - \frac{1}{r} \slashed{q} \partial_u  \ov{0}{\Psi}  + \\ \frac{1}{r^2}\left( - \slashed{q} \partial_u  \ov{1}{\Psi} + \slashed{k} \ov{0}{\Psi} + \partial_a \slashed{q} \partial^a \ov{0}{\Psi} \right) + \cdots
\ee
The vanishing of the $1/r$ term leads to \eqref{asymPsi}, which we  write as\footnote{In \eqref{asymPsi} and \eqref{Psi0itou}  we are setting to zero a possible  $u$-independent spinor that may not be in the kernel of  $\slashed{q}$. It may be that such spinor is needed in an extended version of the radiative phase space.} 
\be \label{Psi0itou}
\ov{0}{\Psi} = \psi_+ \ubold_+ + \psi_- \ubold_-
\ee
where 
\be \label{spinorsubold}
\ubold_+ = \frac{1}{2^{1/4}}  \begin{pmatrix} 0 \\ 0 \\ 1 \\ z   \end{pmatrix} , \quad  \ubold_-  = \frac{1}{2^{1/4}} \begin{pmatrix} -\zb  \\ 1 \\ 0 \\ 0  \end{pmatrix} ,
\ee
span the kernel of the matrix $\slashed{q}$.  

$\ov{1}{\Psi}$ is to be determined from  the vanishing of the $1/r^2$ term in \eqref{Diraceqinr}.  Due to the non-invertibility of $\slashed{q}$, it is not immediately obvious this equation can be solved. However, using that $k^\mu= 1/2 \partial^a \partial_a q^\mu$, the equation can be brought into the form
\be
0  =  - \slashed{q}( \partial_u  \ov{1}{\Psi} + \frac{1}{2} \partial^a \partial_a \ov{0}{\Psi}) + \frac{1}{2}  \partial^a \partial_a ( \slashed{q} \ov{0}{\Psi} ) .
\ee
Since the last term vanishes due to \eqref{Psi0itou}, the equation fixes $\ov{1}{\Psi}$ modulo elements in the kernel of $\slashed{q}$:\footnote{The first term in \eqref{Psi1} is what one gets from solving  $\square \Psi=0$. The apparent indeterminacy in $\ov{1}{\Psi}$ gets fixed upon requiring integrability on the equation for $\ov{2}{\Psi}$. However, we do not need such terms for the present discussion.} 
\ba \label{Psi1}
\ov{1}{\Psi} &= & - \frac{1}{2}\partial_u^{-1} \partial^a \partial_{a}  \ov{0}{\Psi} + \text{ker}(\slashed{q}). \\
&= & - \sum_s   \partial_u^{-1}  \partial_a  \psi_s  \partial^{a} \ubold_s  + \text{ker}(\slashed{q}).
\ea
Substituting in \eqref{Ja} we  get
\ba
j_z & =& -e\left( \psib_- \partial_z \partial_u^{-1} \psi_- + \partial_z \partial_u^{-1} \psib_+ \psi_+  \right) ,\\
j_{\zb} & =& -e\left(  \psib_+ \partial_{\zb} \partial_u^{-1} \psi_+ + \partial_{\zb} \partial_u^{-1} \psib_- \psi_- \right) ,
\ea
where we used that $\bar{\ubold}_{s} \partial_a \slashed{q}  \ubold_{s'}=0$ (which ensures $j_a$  is independent of the $\text{ker}(\slashed{q})$ indeterminacy in $\ov{1}{\Psi}$), together with the fact that the only nonzero components of $\bar{\ubold}_{s} \partial_a \slashed{q} \partial_b \ubold_{s'}$ are
\be
\bar{\ubold}_{+} \partial_{\zb} \slashed{q} \partial_z \ubold_{+}= \bar{\ubold}_{-} \partial_{z} \slashed{q} \partial_{\zb} \ubold_{-}=1.
\ee
Defining $\delta^{\lps}_Y = \{ \cdot, Q^\lps_Y\}$ with the PBs \eqref{elemPBs} one obtains
\be \label{delLPS}
\begin{aligned}
\delta^{\lps}_Y \psi_+ & =  i e \left( Y^a \partial_a  + u \partial_a Y^a \partial_u + \partial_z Y^z\right) \partial_u^{-1} \psi_+,\\
\delta^{\lps}_Y \psi_- & =  i e \left( Y^a \partial_a  + u \partial_a Y^a \partial_u + \partial_{\zb} Y^{\zb} \right) \partial_u^{-1} \psi_-, \\
\delta^{\lps}_Y \dot A_z & =  - \partial_z^2 Y^z .
\end{aligned}
\ee

\subsection{Naive commutator of two fermionic generators} 
In this section we evaluate the (anti) commutator between two fermionic variations $\delta_\chi$ defined in \eqref{delchiradPB}. As discussed there, one should not take this result too seriously, since the expressions do not incorporate ``corner'' fermionic degrees of freedom that are presumably needed to have a well defined phase space action. 

To facilitate the computation, let us introduce the notation
\be
\chi_+ = \chi_{z - }, \quad \chi_- = \chi_{\zb +}
\ee 
so that \eqref{delchiradPB} can be written more compactly as,
\be \label{delchirads}
\begin{aligned}
\delta_\chi \psi_s &= D_s \chi_{-s},  \\
\delta_\chi \dot A_s & = \frac{e}{2}\left( \chib_{-s} \psi_s - \chi_s \psib_{-s} \right).
\end{aligned}
\ee
We focus on the anticommutator $[\delta_\chi, \delta_{\chi} ] = 2 \delta_\chi \delta_{\chi}$, since the general case can be obtained from this basic one. Acting twice with \eqref{delchirads} one finds,
\be \label{delchidelchi}
\begin{aligned}
[ \delta_\chi, \delta_{\chi} ]  \psi_+ & = -i e^2 \left( \chib_- \chi_-  \partial_u^{-1} \psi_+ -\chi_+ \chi_- \partial^{-1}_u \psib_- \right)  \\
 [ \delta_\chi, \delta_{\chi} ] \dot{A}_z & = -e \left(D_z \chib_+ \chi_+  +  \chib_- D_z \chi_-  \right) 
\end{aligned}
\ee
(the corresponding expressions for $\psi_-$ and $\dot A_{\zb}$ can be obtained from \eqref{delchidelchi} by interchanging $+ \leftrightarrow -$ and $z \leftrightarrow \zb$). The result bears certain resemblance with  $\delta^\lps_Y$ \eqref{delLPS}  if 
\be
Y^z \sim e \partial_z^{-1} \chib_s  \chi_s.
\ee

\end{document}